\def\half{\frac{1}{2}}
\def\nll{\hfil\hfil\linebreak\noindent} 
\def\hi#1#2{$#1$\kern -2pt-#2} 
\def\hy#1#2{#1-\kern -2pt$#2$} 
\def\dbox#1{\hbox{\vrule 
\vbox{\hrule \vskip #1\hbox{\hskip #1\vbox{\hsize=#1}\hskip #1}\vskip #1 
\hrule}\vrule}}  
\def\qed{\begin{flushright}~{\dbox{0.05true in}}\end{flushright}} 

\documentclass[ams]{revtex4}
\usepackage{graphicx}
\begin{document}
\hspace*{4.5 in} CUQM - 127
\vspace*{0.4 in}

\title{Special comparison theorem for the Dirac equation}
\author{Richard L. Hall}
\address{Department of Mathematics and Statistics, Concordia University,
1455 de Maisonneuve Boulevard West, Montreal,
Quebec, Canada H3G 1M8}
\email{rhall@mathstat.concordia.ca}
\begin{abstract}If a central vector potential $V(r,a)$ in the Dirac equation is monotone in a parameter $a,$ then a discrete eigenvalue $E(a)$ is monotone in $a.$  For such a special class of comparisons, this generalizes an earlier comparison theorem that was restricted to node free states. Moreover, the present theorem applies to every discrete eigenvalue.
\end{abstract}
\pacs{03.65.Ge, 03.65.Pm}
\keywords{Dirac equation, discrete spectrum, comparison theorem}
\vskip0.2in
\maketitle
\section{Introduction and main result}
We consider a single particle that is bound by an attractive central vector potential $V$ in $d\ge 1$ spatial dimensions and obeys the Dirac equation.  For a central potential in $d$ dimensions the Dirac equation can be written \cite{jiang} in natural units $\hbar=c=1$ as
\begin{equation}\label{eq1}
i{{\partial \Psi}\over{\partial t}} =H\Psi,\quad {\rm where}\quad  H=\sum_{s=1}^{d}{\alpha_{s}p_{s}} + m\beta+V,
\end{equation}
$m$ is the mass of the particle, $V$ is a spherically symmetric vector potential, and $\{\alpha_{s}\}$ and $\beta$  are the Dirac matrices which satisfy anti-commutation relations; the identity matrix is implied after the potential $V$. For stationary states, some algebraic calculations in a suitable basis lead to a pair of first-order linear differential equations in two radial functions $\{\psi_1(r), \psi_2(r)\}$, where $r = ||\mathbf{r}||.$  For $d > 1,$ these functions vanish at $r = 0$, and, for bound states, they may be normalized by the relation 
\begin{equation}\label{eq2}
(\psi_1,\psi_1) + (\psi_2,\psi_2) = \int\limits_0^{\infty}(\psi_1^2(r) + \psi_2^2(r))dr = 1.
\end{equation}
We use inner products {\it without} the radial measure $r^{(d-1)}$ because the factor $r^{\frac{(d-1)}{2}}$ is already built in to each radial function. These radial functions satisfy the coupled equations
\begin{eqnarray}
E\psi_1 &=& (V+m)\psi_1 + (-\partial + k_{d}/r)\psi_2\label{eq3}\\
E\psi_2 &=& (\partial + k_{d}/r)\psi_1 + (V-m)\psi_2\label{eq4},
\end{eqnarray}
where $k_1 = 0,$ $k_{d}=\tau(j+{{d-2}\over{2}}),~d >1$,  and $\tau = \pm 1.$  We note that the variable $\tau$ is sometimes written $\omega$, as, for example in the book by Messiah \cite{messiah}, and the radial functions are often written $\psi_1 = G$ and $\psi_2 = F,$ as in the book by Greiner \cite{greiner}.  We shall assume that the potential $V$ is such that there is a discrete eigenvalue $E$ and that Eqs.(\ref{eq3},~\ref{eq4}) are the eigenequations for the corresponding radial eigenstates.  In this paper we shall present the problem explicitly for the cases $d > 1.$ The same arguments go through {\it mutatis mutandis} for the case $d=1$: in this case $k_1 = 0,$ the states can be classified as even or odd, and the normalization (\ref{eq2}) becomes $\int_{-\infty}^{\infty}\left(\psi_1^2(x) + \psi_2^2(x)\right)dx = 1.$   

We are interested in the relation that is induced between the eigenvalues associated with two distinct potentials $V^{(1)}$ and $V^{(2)}$ if it is known that the potentials, as functions, are ordered $V^{(1)} \leq V^{(2)}.$  The well-known comparison theorem of non-relativistic quantum mechanics states that $V^{(1)} \leq V^{(2)} \Rightarrow 
E^{(1)} \leq E^{(2)}.$  Since the eigenvalues in the Schr\"odinger case may be characterized variationally, the proof is almost immediate.  However, the Dirac energy operator is not bounded below and the variational approach is not readily applicable.  We have already solved this problem for the Dirac ground state in $d = 3$ provided that it is node free: under these conditions we proved \cite{halld} that  $V^{(1)} \leq V^{(2)}~\Rightarrow~E^{(1)}_0 \leq E^{(2)}_0.$ Subsequently this result was extended by Chen \cite{chen} to all dimensions $d\geq 1.$  The purpose of the present paper is to derive a much stronger Dirac result of this type, but valid for a restricted class of comparison potentials.  We suppose that the attractive central potential $V(r,a)$ depends smoothly on a parameter $a$. This in turn implies that each discrete eigenvalue $E(a)$, and the corresponding radial functions $\psi_1(r,a)$ and $\psi_2(r,a)$, depend on $a.$  For each eigenvalue we establish the following
\medskip\nll{\bf Theorem 1}
\begin{equation}\label{eq5}
\partial V/\partial a \geq 0~~\Rightarrow~~E'(a) \geq 0~~~{\rm and}~~~\partial V/\partial a \leq 0~~\Rightarrow~~E'(a) \leq 0.\nonumber
\end{equation} 
Examples of such potentials are (i) $V(r,a) = a f(r),$ where $a >0$ is a coupling parameter and the potential shape $f(r)$ does not change sign, and (ii) $V(r,a) = -\alpha/(r + a),$ $\alpha > 0,$ which satisfies $\partial V/\partial a > 0$ and  $\partial V/\partial \alpha < 0$.
\section{Proof of Theorem 1}
By differentiating the normalization integral (\ref{eq2}) partially with respect to $a$, and writing $\partial \psi/\partial a = \psi_a$ for each wave-function component, we obtain the orthogonality relation 
\begin{equation}\label{eq6}
(\psi_{1a},\psi_{1}) + (\psi_{2a},\psi_{2}) = 0.
\end{equation}
We now differentiate Eqs.(\ref{eq3},~\ref{eq4}) with respect to $a$ to give
\begin{eqnarray}
E'(a)\psi_1 + E(a)\psi_{1a} &=& V_a\psi_1 + (V+m)\psi_{1a} + (-\partial + k_{d}/r)\psi_{2a}\label{eq7}\\
E'(a)\psi_2 + E(a)\psi_{2a} &=& (\partial + k_{d}/r)\psi_{1a} + V_a\psi_2 + (V-m)\psi_{2a}\label{eq8}.
\end{eqnarray}
If we integrate the linear combination ${\rm (7)}\psi_1 + {\rm (8)} \psi_2$ of  Eqs.(\ref{eq7},~\ref{eq8}) on $[0,\infty)$ we obtain
\begin{equation}\label{eq9}
E'(a)[(\psi_1,\psi_1) + (\psi_2,\psi_2)] = [(\psi_1,V_a\psi_1) + (\psi_2,V_a\psi_2)] + W,
\end{equation}
where
\begin{eqnarray}\label{eq10}
W & = &(\psi_{1a},(V+m),\psi_1) + (\psi_1, (-\partial + k_d/r)\psi_{2a}) - E(a)(\psi_{1a},\psi_1)\nonumber\\
& + & (\psi_{2a},(V-m),\psi_2) + (\psi_2, (\partial + k_d/r)\psi_{1a}) - E(a)(\psi_{2a},\psi_2).
\end{eqnarray}
We now show that $W = 0.$ We first need to establish the anti-symmetric relation
\begin{equation}\label{eq11}
(\psi,\partial\phi) = - (\partial \psi, \phi),
\end{equation}
where $\partial$ represents the differential operator $\partial = \partial/\partial r.$ This is achieved by the use of an integration by parts and the boundary conditions.
In $d > 1$ dimensions we have
\begin{equation}\label{eq12}
(\psi,\partial\phi) = \left[\psi(r)\phi(r)\right]_{0}^{\infty} -\int\limits_{0}^{\infty}(\partial\psi(r))\phi(r)\,dr = - (\partial \psi, \phi).
\end{equation}
We note parenthetically that in $d=1$  dimension we obtain similarly 
\begin{equation}\label{eq13}
(\psi,\partial_x\phi) = \left[\psi(x)\phi(x)\right]_{-\infty}^{\infty} -\int\limits_{-\infty}^{\infty}(\partial_x\psi(x))\phi(x)\,dx = - (\partial_x \psi, \phi).
\end{equation}
In view of Eq.(\ref{eq11}) we may therefore rewrite $W$ (for $d > 1$) as
\begin{eqnarray}\label{eq14}
W &=& \left(\psi_{1a},[(V+m)\psi_1+(-\partial + k_d/r)\psi_2 - E(a)]\psi_1\right)\nonumber\\
&+& \left(\psi_{2a},[(V-m)\psi_2+(\partial + k_d/r)\psi_1 - E(a)]\psi_2\right).
\end{eqnarray}
From the eigen equations Eqs.(\ref{eq3},~\ref{eq4}) we conclude that $W = 0.$  The normalization 
Eq.(\ref{eq2}) together with Eq.(\ref{eq9}) imply
\begin{equation}\label{eq15}
E'(a) = (\psi_1,V_a\psi_1) + (\psi_2,V_a\psi_2).
\end{equation}
Thus, if $V_a$ does not change sign, nor does $E'(a),$ and the signs are the same.  This completes the proof of the theorem.\qed

\section{Conclusion}
The principal result of this paper should be regarded more as a contribution to the qualitative theory of Dirac spectra than to approximation theory. It confirms experience with explicit potentials such as the Coulomb potential $V = -\alpha/r$ for which the known exact spectral function
$$E(\alpha) = \left[1 + \frac{\alpha^2}{\left[n -j-\half + \sqrt{(j+\half)^2 -\alpha^2}\right]^2}\right]^{-\half}$$
satisfies $E'(\alpha) < 0$ for every eigenvalue.  In order to support the application of envelope methods to screened-Coulomb problems discussed in Ref.\cite{halld} we would still need the more general theorem of that paper.  However, for a cutoff Coulomb potential such as $V(r,\alpha,a) = -\alpha/(r+a)$, since $\partial V/\partial a > 0,$ we know immediately from Theorem 1 that $\partial E(\alpha,a)/\partial a > 0$ for each discrete Dirac eigenvalue.  Consequently, for example, it is now clear that the complicated non-monotonic behavior \cite{barton, hallkg} of the corresponding Klein-Gordon spectrum generated by a cutoff Coulomb potential will not occur under the Dirac equation: Theorem 1 tells us that every discrete Dirac eigenvalue decreases monotonically with $\alpha$ and increases monotonically with $a.$  It is thus possible for us to establish such monotone properties even though it may be difficult to solve a given problem exactly or to characterize the Dirac spectrum variationally. 

 \section*{Acknowledgment}
Partial financial support of his research under Grant No.~GP3438 from~the Natural
Sciences and Engineering Research Council of Canada is gratefully acknowledged. 


\begin{thebibliography}{99}
\bibitem{jiang}Y. Jiang, J. Phys. A {\bf 38} 1157 (2005).
\bibitem{messiah}A. Messiah, {\it Quantum Mechanics}, (North Holland, Amsterdam, 1962). The Dirac equation for central fields is discussed on page 928.
\bibitem{greiner}W. Greiner {\it Relativistic Quantum Mechanics}, (Springer, Heidelberg, 1990). The Dirac equation for the Coulomb central potential is discussed on page 178.
\bibitem{halld}R. L. Hall, Phys. Rev. Lett. {\bf 83}, 468 (1999).
\bibitem{chen}G. Chen, Phys. Rev A {\bf 72}, 044102 (2005).
\bibitem{barton}G. Barton, J. Phys. A {\bf 40}, 1011 (2007).
\bibitem{hallkg}R. L. Hall, Phys. Lett. A {\bf 372}, 12 (2007).


\end{thebibliography}
\end{document}